\documentclass[11pt,oneside]{amsart}

\usepackage{epsfig,psfrag,subfigure}
\usepackage{epstopdf}
\usepackage{amsmath,amssymb,amsfonts,latexsym, mathrsfs,euscript}
\usepackage{graphicx,graphics,cite}
\usepackage{wrapfig}
\usepackage[usenames]{color}
\usepackage{geometry}                % See geometry.pdf to learn the layout options. There are lots.
\usepackage{colortbl,booktabs,ctable,multirow}
\usepackage{epstopdf}
\usepackage{subfig}

\DeclareGraphicsExtensions{.pdf,.png,.jpg}

\newcommand{\beq}{\begin{equation}}
\newcommand{\eeq}{\end{equation}}
\newcommand{\beqa}{\begin{eqnarray}}

\newcommand{\eeqa}{\end{eqnarray}}

\newcounter{nfig}

\newcommand{\ignore}[1]{}

\newcommand\ds{\displaystyle}
\newcommand\half{\displaystyle\frac{1}{2}}
\newcommand\bs{\boldsymbol}
\title[SW and EPDiff equations]{A connection between the shallow-water equations and the Euler-Poincar\'e equations}

\author{Roberto Camassa}
\address{Department of Mathematics, University of North Carolina, Chapel Hill, 27599, USA}
\email{camassa@amath.unc.edu}

\author{Long Lee}
\address{Department of Mathematics, University of Wyoming, Laramie, WY 82071-3036, USA}
\email{llee@uwyo.edu}

\date{April 18, 2014}

\begin{document}

\maketitle

\begin{abstract}
The Euler-Poincar\'e differential (EPDiff) equations and the shallow water (SW) equations share similar wave characteristics. Using the Hamiltonian structure of the SW equations with flat bottom topography, we establish a connection between the EPDiff equations and the SW equations in one and multi-dimensions. Additionally, we show that the EPDiff equations can be recast in a curl formulation. 
\end{abstract}

\begin{description}
\item[{\footnotesize\bf keywords:} ]
{\footnotesize Euler-Poincar\'e differential  equations, shallow-water equations, flat bottom topography, Hamiltonian.}
\end{description}

\section{Introduction}
The Euler-Poincar\'e differential (EPDiff) equations are a class of models for ideal incompressible fluids in three dimensions \cite{bib:hmr98}.  Holm and Staley show that numerical simulations for the EPDiff equations exhibit similar nonlinear behavior of wavefront reconnections as that observed for internal waves in the South China Sea \cite{bib:hs13}. Camassa et al. show that the EPDiff equations exist traveling-wave solutions, and the dynamics of the two-dimensional (2-D) solitons are similar to that of the one-dimensional (1-D) peakons of the shallow-water wave equation (also known as the Camassa-Holm equation). Similar to the 1-D Camassa-Holm equation, the solitary waves for this class of equations can be made to correspond to interacting particles of a finite-degree-of-freedom Hamiltonian system, which we refer to as the $N$-soliton (particle) system.  The power of the elliptic operator associated with the EPDiff equations plays an important role for the soliton dynamics \cite{NP, bib:ckl14}, since the solitons are the Green's functions of the elliptic operator associated with the EPDiff equations. For the special case of a fractional power $\nu=3/2$, the Green's function becomes a 2-D extension of the peakon solution of the Camassa-Holm equation. In particular, when the motion of the 2-D solitons is confined in a straight line, the reduced $N$-soliton system of the EPDiff equations enjoys the same properties of complete integrability as that of the 1-D Camassa-Holm equation \cite{bib:ckl14}. Because the wave characteristics modeled by the EPDiff equations are similar to that of the shallow-water models, this motivates us to explore the connection between the shallow-water (SW) equations and the EPDiff equations.

Camassa et al. establish the connection between a SW model, also referred to as the great lake equations, and the Green-Naghdi equations \cite{bib:chl96}. This letter will follow the similar strategy as that in \cite{bib:chl96} to establish the connection between the SW equations with flat bottom topography and the EPDiff equations. 
  
\section{One-dimensional space}
We first illustrate the connection in 1-D space and then we extend the case to multi-dimensional spaces in the next section. The one-dimensional SW equations with flat bottom topography are written as 
\begin{equation}\label{eq:1D_SW_Airy}
\begin{split}
&u_t+uu_x+g\eta_x=0\\
&\eta_t+(\eta u)_x=0,
\end{split}
\end{equation}
where $u(x, t)$ is the velocity, $\eta(x, t)$ is the vertical displacement of free surface, and $g$ is the gravity.  We introduce a variable, $m \equiv \eta u$, as a function of the variables of the SW equations. The Hamiltonian for the SW equations is then defined as
\begin{equation}\label{eq:H}
\mathcal{H}=\int_{-\infty}^{\infty}\left(\displaystyle\frac{(\eta u)u}{2}+\displaystyle\frac{g\eta^2}{2}\right)dx=\int_{-\infty}^{\infty}\left(\displaystyle\frac{m u}{2}+\displaystyle\frac{g\eta^2}{2}\right)dx.
\end{equation}
With the Hamiltonian formula, the SW equations is equivalent to
\begin{equation}\label{eq:1D_Hamiltonian}
\left(
\begin{array}{c}
m_t\\
\eta_t
\end{array}
\right)
=-\left(
\begin{array}{cc}
m\partial_x+(\partial_x) m & \eta\partial_x\\
\partial_x\eta& 0
\end{array}
\right)
\left(
\begin{array}{c}
\displaystyle\frac{\delta\mathcal{H}}{\delta m}\\ 
\displaystyle\frac{\delta\mathcal{H}}{\delta\eta}
\end{array}
\right),
\end{equation}
where $\displaystyle\frac{\delta\mathcal{H}}{\delta m}$ and $\displaystyle\frac{\delta\mathcal{H}}{\delta\eta}$ are the variational derivatives of $\mathcal{H}$.

To check that Eq. (\ref{eq:1D_Hamiltonian}) is equivalent to the SW equations (\ref{eq:1D_SW_Airy}), we compute
\begin{equation}\label{eq:1D_dela_H}
\begin{split}
\displaystyle\frac{\delta\mathcal{H}}{\delta m}&=\displaystyle\frac{\delta}{\delta m}\left(\displaystyle\frac{m^2}{2\eta}+\displaystyle\frac{g\eta^2}{2}\right)=\displaystyle\frac{m}{\eta}=u,\\
\displaystyle\frac{\delta\mathcal{H}}{\delta\eta}&=\displaystyle\frac{\partial}{\partial\eta}\left(\displaystyle\frac{m^2}{2\eta}+\displaystyle\frac{g\eta^2}{2}\right)=-\displaystyle\frac{m^2}{2\eta^2}+g\eta=-\frac{1}{2}u^2+g\eta.
\end{split}
\end{equation}
Hence the first equation (momentum equation) in Eq. (\ref{eq:1D_Hamiltonian}) is 
\begin{equation}\label{eq:m_t}
\begin{split}
m_t&=-\left(m\partial_x+(\partial_x) m\right)u-\eta\partial_x\left(\displaystyle-\frac{1}{2}u^2+g\eta\right)\\
       &=-mu_x-(mu)_x+\eta uu_x-g\eta\eta_x\\
       &=-(mu)_x-g\eta\eta_x,    
\end{split}
\end{equation}
and the second equation in Eq. (\ref{eq:1D_Hamiltonian}) becomes
\begin{equation}\label{eq:eta_t}
\eta_t=-\partial_x(\eta u),   
\end{equation}
which is the equation of conservation of mass.
Note that Eq. (\ref{eq:m_t}) is equivalent to the first equation (momentum equation) in Eq. (\ref{eq:1D_SW_Airy}). To see this, multiplying $\eta$ to the first equation in Eq. (\ref{eq:1D_SW_Airy}) yields
\begin{equation}\label{eq:eta_u}
\eta(u_t+uu_x+g\eta_x)=(\eta u)_t-\eta_t u+\eta uu_x+g\eta\eta_x=0.
\end{equation}
From Eq. (\ref{eq:eta_t}), we have 
\begin{equation}
\eta_t=-(\eta u)_x=-m_x,
\end{equation}
and then Eq. (\ref{eq:eta_u}) becomes
\begin{equation}\label{eq:m}
m_t+um_x+mu_x+g\eta\eta_x=0,
\end{equation}
or
\begin{equation}\label{eq:m2}
m_t+(mu)_x+g\eta\eta_x=0,
\end{equation}
which is identical to Eq. (\ref{eq:m_t}). Now suppose we neglect the vertical displacement of the free surface,  i.e. in Eq. (\ref{eq:1D_Hamiltonian}), $\eta=0$ in the first equation of the Hamiltonian operation, then we obtain the 1-D EPDiff equations 
\beq\label{eq:1D_EP}
m_t+(mu)_x+mu_x =0,
\eeq
where the momentum-like variable $m$ can be related to the velocity $u$ by a self-adjoint elliptic operator 
\beq
m=(1-\alpha^2\partial_{xx})^{\nu}u.
\eeq 

%Hence we establish the connection between the one-dimensional SW equations with the 

\section{Multi-dimensional spaces}

The multi-dimensional SW equations with flat bottom topography are written as 
\begin{equation}\label{eq:2D_SW_Airy}
\begin{split}
&\bs{u}_t+(\bs{u}\cdot\nabla)\bs{u}+g\nabla\eta=0\\
&\eta_t+\nabla\cdot(\eta\bs{u})=0,
\end{split}
\end{equation}
where $\bs{u}(x, y, t)=(u, v)^T$ is the velocities, $\eta(x, y, t)$ is the vertical displacement of free surface, and $g$ is the gravity. Similar to the 1-D case, letting $\bs{m}=\eta\bs{u}$,  Eq. (\ref{eq:2D_SW_Airy}) can be written in terms of the Hamiltonian structure 
\begin{equation}\label{eq:2D_Hamiltonian}
\left(
\begin{array}{c}
\dot{m_i}\\
\dot{\eta}
\end{array}
\right)
=-\left(
\begin{array}{cc}
m_j\partial_i+(\partial_j) m_i & \eta\partial_i\\
\partial_j\eta & 0
\end{array}
\right)
\left(
\begin{array}{c}
\displaystyle\frac{\delta\mathcal{H}}{\delta m_j}\\ 
\displaystyle\frac{\delta\mathcal{H}}{\delta\eta}
\end{array}
\right),
\end{equation}
where the Hamiltonian is defined as
\begin{equation}
\mathcal{H}=\half\int_{-\infty}^{\infty}dx\int_{-\infty}^{\infty}dy\left(\bs{m}\cdot\bs{u}+g\eta^2\right),
\end{equation}
where 
\begin{equation}
\bs{m}\cdot\bs{u}=\ds\frac{|m|^2}{\eta}.
\end{equation}
To check that Eq. (\ref{eq:2D_Hamiltonian}) is equivalent to Eq. (\ref{eq:2D_SW_Airy}), we compute
\begin{equation}\label{eq:2D_dela_H}
\begin{split}
\displaystyle\frac{\delta\mathcal{H}}{\delta m_j}&=u_j,\\
\displaystyle\frac{\delta\mathcal{H}}{\delta\eta}&=\displaystyle\frac{\partial}{\partial\eta}\left(\displaystyle\frac{|m|^2}{2\eta}+\displaystyle\frac{g\eta^2}{2}\right)=-\displaystyle\frac{|m|^2}{2\eta^2}+g\eta=-\frac{1}{2}\bs{u}\cdot\bs{u}+g\eta.
\end{split}
\end{equation}
Eq. (\ref{eq:2D_dela_H}) implies that the first equation (momentum equation) in Eq. (\ref{eq:2D_Hamiltonian}) has the form
\begin{equation}\label{eq:m_dot}
\begin{split}
\dot{m_i}&=-\left(m_j \partial_iu_j+\partial_j(m_iu_j)-\eta\partial_i\ds\frac{u_ju_j}{2}+g\eta\partial_i\eta\right)\\
               &=-m_j\partial_iu_j-\partial_j(m_iu_j)+\eta u_j\partial_iu_j - g\eta\partial_i \eta\\
               &=-\partial_j(m_iu_j)-g\eta\partial_i\eta.
\end{split}
\end{equation}
Similarly, the second equation in Eq. (\ref{eq:2D_Hamiltonian}) is
\begin{equation}
\dot{\eta} = -\partial_j(\eta u_j),
\end{equation}
or 
\begin{equation}\label{eq:eta_dot}
\eta_t + \nabla\cdot(\eta\bs{u}) = 0.
\end{equation}
We now show that Eq. (\ref{eq:m_dot}) is equivalent to the first equation in (\ref{eq:2D_SW_Airy}). Multiplying $\eta$ to the first equation of Eq. (\ref{eq:2D_SW_Airy}), we obtain 
\begin{equation}
\eta\left(\bs{u}_t+(\bs{u}\cdot\nabla)\bs{u}+g\nabla\eta\right)=0,
\end{equation}
or
\begin{equation}
(\eta\bs{u})_t-\eta_t\bs{u}+(\bs{m}\cdot\nabla)\bs{u}+g\eta\nabla\eta=0.
\end{equation}
Using the definition of $\bs{m}$, we obtain
\begin{equation}\label{eq:m_dot_vec}
\bs{m}_t+(\nabla\cdot\bs{m})\bs{u}+(\bs{m}\cdot\nabla)\bs{u}+g\eta\nabla\eta=0,
\end{equation}
since $\eta_t=-\nabla\cdot\bs{m}$ by equation (\ref{eq:eta_dot}).

We write Eq. (\ref{eq:m_dot_vec}) in its tensor form:
\begin{equation}
\dot{m_i} + (\partial_j m_j)u_i+ m_j\partial_j u_i + g\eta\partial_i\eta=0,
\end{equation}
or
\begin{equation}\label{eq:m_dot_1}
\dot{m_i}+\partial_j(m_j u_i) +g\eta\partial_i\eta = 0.
\end{equation}
Now, since $m_j=\eta u_j$, which implies $m_j u_i = u_j m_i$, Eq. (\ref{eq:m_dot_1}) is equivalent to
\begin{equation}
\dot{m_i}+\partial_j(m_i u_j) +g\eta\partial_i\eta = 0.
\end{equation} 
The above equation is identical to Eq. (\ref{eq:m_dot}).

Similar to the 1-D example, if we ignore the vertical displacement of the free surface, i.e. $\eta=0$ in  the first equation of the Hamiltonian operation, we have
\begin{equation}
\begin{split}
\dot{m_i}&=-m_j\partial_i u_j - \partial_j(m_i u_j)\\
                &=-m_j\partial_i u_j - m_i\partial_j u_j - u_j\partial_j m_i,
                \end{split}
\end{equation}
or in short-hand vector notation
\begin{equation}\label{eq:SW_momentum}
\bs{m}_t +(\nabla\bs{u})^{T}\bs{m} + (\bs{u}\cdot\nabla)\bs{m} +\bs{m}(\nabla\cdot\bs{u}) = 0.
\end{equation}
The notation $(\nabla\bs{u})^{T}$ denotes the transpose of the matrix $\nabla\bs{u}$.

\subsection{The Euler-Poincar\'e differential equations}

Eq. (\ref{eq:SW_momentum}) is referred to as the Euler-Poincar\'e differential (EPDiff) equations. Conventionally, the velocity $\bs{u}$ and the momentum variable $\bs{m}$ are formally related by
\begin{equation}\label{eq:Yukawa}
\bs{m}=\mathcal{L}\bs{u}.
\end{equation}
With appropriate boundary conditions, the operator $\mathcal{L}$ is assumed to be invertible, with its inverse being explicitly written in terms of the corresponding Green function $\bs{G}$, so that 
\begin{equation}\label{eq:convol}
\bs{u} =\bs{G}*\bs{m}.
\end{equation}
A particular choice of $\mathcal{L}$ is $\mathcal{L}=\mathcal{L}^{\nu}$, the Yukawa operator, defined by
\begin{equation}\label{eq:elliptic}
\mathcal{L}^{\nu}=(\bs{I}-\alpha^2\nabla^2)^{\nu},
\end{equation}
parameterized by $\alpha^2\le 1$ and $\nu >0$. For any $\nu >0$, including non-integer values. Eq. (\ref{eq:Yukawa}) is defined in the Fourier space.
\beq\label{eq:m-Fourier}
\bs{\hat{u}} = (\hat{\mathcal{L}^{\nu}})^{-1}\bs{\hat{m}},\quad\text{where}\,\,\,(\hat{\mathcal{L}^{\nu}})^{-1} = \frac{1}{(1+\alpha^2 k^2)^{\nu}},\quad k=\sqrt{k_1^2+k_2^2\cdots+k_n^2},
\eeq
where $k_i$ is the the wavenumber of $i^{th}$ component. Since $\mathcal{L}^{\nu}$ is rotationally invariant and diagonal, $\bs{G}(\bs{x}) = G_{\nu-n/2}(|\bs{x}|)\bs{I}$ for a scalar function $G_{\nu-n/2}$, where $|\bs{x}|=\sqrt{x_1^2+x_2^2+\cdot+x_n^2}$.   The scalar Green's function $G_{\nu-n/2}$ is a function combining the modified Bessel function of the second kind and the Gamma function  
%For $\bs{x}\in\mathbb{R}^{n}$, the Green's function of $\mathcal{L}=(\bs{I}-\alpha^2\nabla^2)^{\nu}$ is 
\beq\label{eq:Green-nD}
G_{\nu-n/2}(|\bs{x}|) = \frac{2^{n/2-\nu}}{(2\pi\alpha)^{n/2}\alpha^{\nu}\Gamma(\nu)}|\bs{x}|^{\nu -n/2}K_{\nu -n/2}\left(\frac{|\bs{x}|}{\alpha}\right),
\eeq
where $K_{\nu -1}$ is the modified Bessel function of the second kind of order $\nu -n/2$ and $\Gamma(\nu)$ is the Gamma function \cite{bib:mumford}.

Eq. (\ref{eq:SW_momentum}) can be recast in a curl formulation
\begin{equation}\label{eq:EP_curl}
\bs{m}_t - \bs{u}\times\text{curl}\,\bs{m}+\nabla(\bs{u}\cdot\bs{m})+\bs{m}(\nabla\cdot\bs{u}) = 0.
\end{equation}
We now show that Eq. (\ref{eq:EP_curl}) is the same as  Eq. (\ref{eq:SW_momentum}). Note that the second term in Eq. (\ref{eq:EP_curl}) is
\begin{equation}
\begin{split}
-\bs{u}\times\text{curl}\,\bs{m}&=-\varepsilon_{ijk}u_j(\text{curl}\,m)_k\\
                                                    &=-\varepsilon_{ijk}u_j\varepsilon_{krs}\partial_r m_s\\
                                                    &=-(\delta_{ir}\delta_{js}-\delta_{is}\delta_{jr}) u_j\partial_r m_s\\
                                                    &=-(u_j\partial_i) m_j+(u_j\partial_j) m_i.
\end{split}
\end{equation}
Hence Eq. (\ref{eq:EP_curl}) is equivalent to
\beq
\dot{m}_i - u_j\partial_i m_j + (u_j\partial_j) m_i + \partial_i(m_j u_j) + m_i\partial_j u_j = 0
\eeq
or
\beq
\dot{m}_i  - u_j\partial_i m_j + (u_j\partial_j) m_i + m_j\partial_i u_j + u_j\partial_i m_j  + m_i\partial_j u_j = 0,
\eeq
or
\beq\label{eq:tensor}
\dot{m}_i + m_j\partial_i u_j + (u_j\partial_j) m_i + m_i\partial_j u_j = 0.
\eeq
In vector notation, Eq. (\ref{eq:tensor}) is 
\begin{equation}
\begin{split}
&\bs{m}_t  + m_j\nabla u_j + (\bs{u}\cdot\nabla)\bs{m} + \bs{m}(\nabla\cdot\bs{u}) \\
=&\bs{m}_t + (\nabla\bs{u})^{T}\,\bs{m} + (\bs{u}\cdot\nabla)\bs{m}  +\bs{m}(\nabla\cdot\bs{u})\\
=&0,
\end{split}
\end{equation}
which is Eq. (\ref{eq:SW_momentum}).

\section{Conclusion}

The EPDiff equations have links with fluid dynamics. In particular, the wave dynamics of the EPDiff equations have the similar characteristics as that of the SW equations. We explore the connection between the EPDiff equations and the SW equations without bottom topography through the Hamiltonian structure of the SW equations. We show that the EPDiff equations are the SW equations without considering the vertical displacement of the free surface.  We also show that the EPDiff equations can be recast in a curl formulation.

%\section{Acknowledgements}
%
%RC acknowledges the support of NSF....

\end{document}